\documentstyle[12pt,epsf,epsfig]{article}
\def\beq{\begin{equation}}
\def\eeq{\end{equation}}
\def\bea{\begin{eqnarray}}
\def\eea{\end{eqnarray}}
\def\bq{\begin{quote}}
\def\eq{\end{quote}}
\def\PL{{ \it Phys. Lett.} }
\def\PRL{{\it Phys. Rev. Lett.} }
\def\NP{{\it Nucl. Phys.} }
\def\PR{{\it Phys. Rev.} }

\def\MPL{{\it Mod. Phys. Lett.} }

\def\gappeq{\mathrel{\rlap {\raise.5ex\hbox{$>$}}
{\lower.5ex\hbox{$\sim$}}}}

\def\lappeq{\mathrel{\rlap{\raise.5ex\hbox{$<$}}
{\lower.5ex\hbox{$\sim$}}}}
\begin{document}
\renewcommand{\theequation}{\arabic{section}.\arabic{equation}}

\pagestyle{empty}
\begin{flushright}
CERN-TH/95-254
\end{flushright}
\vspace*{2cm}
\begin{center}
{\bf  STRING COSMOLOGY: BASIC IDEAS AND GENERAL RESULTS}
 \\
\vspace*{1cm}
 Gabriele Veneziano\\
\vspace*{0.2cm}
Theory Division, CERN\\
1211 CH Geneva 23 \\
\vspace{3cm}
 ABSTRACT
\end{center}

After recalling a few basic concepts from cosmology and
 string theory,
I will outline the main ideas/assumptions underlying
(our own group's approach to) string cosmology and show how
these lead to the definition of a
 two-parameter family of ``minimal" models.
  I will then  briefly explain how to compute, in terms of
those parameters,
 the spectrum of scalar,  tensor and electromagnetic
perturbations, and mention
their most relevant  physical consequences.
More details on the latter part of this talk can be found in
Maurizio Gasperini's contribution to these proceedings.
\vspace*{2cm}
\begin{center}
{\it Talk presented at the} \\
{\it 3rd Colloque Cosmologie, Paris} \\
{\it 7-9 June 1995}
\end{center}
\vspace*{1cm}
\begin{flushleft}
CERN-TH/95-254\\
September 1995
\end{flushleft}





\pagebreak

\section{Basic Facts about Cosmology and Inflation}

It is well known$^{1)}$ that the Standard Cosmological
Model (SCM) works well at
``late" times, its most striking successes being perhaps
 the red shift, the
cosmic microwave background (CMB), and primordial nucleosynthesis.

However, the SCM suffers  from various problems. At the theoretical
level
the most serious of these is the initial singularity problem,
which basically tells us that we cannot have theoretical control
 over the initial conditions of the SCM. At a phenomenological level,
the SCM cannot explain naturally:
\begin{itemize}
\item[i)] the homogeneity and isotropy of our Universe as manifested,
in particular, through the small value of $\Delta T / T = O(10^{-5})$
 observed with COBE$^{2)}$;
\item[ii)] the flatness problem, i.e. the fact that, within an order
of
magnitude,  $\Omega \equiv \rho / \rho_{crit} \sim 1$;
\item[iii)] the origin of large-scale structure.
\end{itemize}

Inflation, i.e. a long phase of accelerated expansion of
the Universe ($\dot a , \ddot a > 0$, where $a$ is
the scale factor),
is the only way known at  present  of solving the above-mentioned
phenomenological problems. Various types of inflationary
models have been proposed [for a review, see 3), 4)]
 each one supposedly mending the problems of the previous version.
Particularly severe are the constraints coming from demanding:

a) a graceful exit with the right amount of reheating;

b) the right amount of large-scale inhomogeneities.

In order to satisfy such constraints, fine-tuned initial
conditions and/or inflaton potentials are necessary.
And this without mentioning the fact that
inflation is not addressing at all the initial singularity problem.

Actually, Kolb and Turner, after reviewing the prescriptions
  for a successful inflation, add$^{4)}$:

``Perhaps the most important -- and most difficult -- task
in building a successful inflationary model is
 to ensure that the inflaton
is an integral part of a sensible model of particle physics.
The inflaton should spring forth from some grander theory
and not vice versa".

I will argue below that superstring theory could be the sought-after
grander theory (what could be better than a theory of
everything?)  naturally providing an inflation-driving
scalar field in the general sense defined again in
 ref. 4):

``It is now apparent that inflation, which was originally so
closely related to Spontaneous Symmetry Breaking,
is a much more general phenomenon....  Stated in its full generality,
inflation involves the dynamical evolution of a very weakly-coupled
scalar field that was originally displaced from the minimum of
its potential."

I hope to convince you that this will be precisely the picture that
we claim  takes place in string cosmology.
In order to substantiate this claim,
I~will have to digress and recall a few basic facts
 in Quantum String Theory.

\section{Basic facts in quantum string theory (QST)}

\setcounter{equation}{0}
I am listing below a few basic properties of strings,
 emphasizing those that are most
relevant for our subsequent discussion.  These are:

{\bf 1.}  Unlike its classical counterpart,
quantum string theory contains a fundamental length scale
 $\lambda_s$.  Such a scale appears in many physical
quantities. It represents, for instance:
\begin{itemize}
\item[a)] Planck's constant (in appropriate units
of energy)~$^{5)}$.
\item[b)] The ultraviolet, short-distance cut-off (equivalently,
a high-momentum cut-off at  $E = M_s \equiv \hbar c / \lambda_s$).
\item[c)] The scale of tree-level masses that are either zero or
$O(M_s)$.  Incidentally,  quantum mechanics
 allows massless strings with
non-zero angular momentum$^{6)}$
while, classically, $M^2 > {\rm const.}
\times J$.  The existence of such states is obviously a crucial
property
of QST,
 without which it could
not pretend to be a candidate theory of all known interactions.
\end{itemize}

{\bf 2.} The absence of free parameters
$(c = \lambda_s = 1$ simply defines
the units of length and  time and are in priciple known numbers
once the centimetre and the second are defined),
 which are replaced by expectation values of fields$^{6)}$.
Basically, some huge (and still largely unknown) symmetry should
 fix the
tree-level ``Lagrangian", while UV finiteness [point 1b)]
preserves full predictive
power at loop level.  Note again here the contrast with
quantum field theory (QFT),
 where, even if the bare
couplings were fixed, the need to renormalize the theory at
loop level would make the
renormalized constants finite but uncalculable.

{\bf 3.} The effective interaction of the massless fields at $E \ll
M_s$ is dictated by the above symmetries and takes the
form of a classical, gauge-plus-gravity field theory with specified
parameters.  It is described by
an effective action$^{7),8)}$ of the (schematic) type:
\bea
\Gamma_{eff} &=& \frac{1}{2} \int d^4x \sqrt{-g}~ e^{-\phi}
\left[\lambda^{-2}_s ({\cal R}
+ \partial_\mu \phi \partial^\mu\phi) + F^2_{\mu\nu} + \bar \psi
D\llap{$/$} \psi +
{\rm higher~derivatives} \right]  \nonumber \\
&& + \left[ {\rm higher~orders~in}~e^\phi \right]~.
\label{31}
\eea
Equation (\ref{31}) contains two dimensionless expansion parameters.
One of them,  $g^2 \equiv e^\phi$, controls the analogue of
  QFT's loop corrections, while the other,
 $\lambda^2 \equiv \lambda_s^2 \cdot \partial^2$, controls
  string-size effects, which are of course absent in QFT.

{\bf 4.} As indicated in (\ref{31}), QST has (actually needs!) a new
particle/field,
the so-called dilaton $\phi$, a scalar massless particle
 (at the perturbative level).  It
appears in $\Gamma_{eff}$ as a Jordan--Brans--Dicke$^{9)}$
 scalar with a ``small" negative
 $\omega_{BD}$ parameter, $\omega_{BD} = -1$.

{\bf 5.} The dilaton's VEV provides$^{8),10)}$ a unified value for:
\begin{itemize}
\item[a)] The gauge coupling(s) at $E = O(M_s)$.
\item[b)] The gravitational coupling in string units.
\item[c)] Yukawa couplings, etc., at the string scale.
\end{itemize}

In formulae:
\bea
\ell^2_p & \equiv & 8 \pi G_N \hbar = e^\phi \lambda^2_s ~, \nonumber
\\
\alpha_{GUT}(\lambda^{-1}_s) &\simeq& \frac{e^\phi}{4\pi}~,
\label{32}
\eea
implying (from $\alpha_{GUT} \approx 1/20$) that the string-length
parameter
$\lambda_s$ will be about 10$^{-32}$~cm.  Note, however, that, in a
cosmological context in which $\phi$ evolves in time, the above
formulae
 can only be taken to give the
$\it{present}$ values of $\alpha$ and $\ell_p/\lambda_s$.  In the
scenario we
will advocate, both quantities were much smaller  in the very
early Universe!

{\bf 6.} Dilaton couplings at large distance are essentially
known$^{11)}$
and can be summarized in the following effective Lagrangian (taking
loop
effects
into account):
\bea
{\cal L}_{eff} &=& \sum_i g_{\phi A_iA_i} \phi A^2_i \nonumber \\
g_{\phi A_iA_i} &=& \sqrt{4\pi G} m^2_i Z_{\phi}^{-1/2} \left. (
\frac{\partial}{\partial \phi} ln \frac{ m^2_i}{M_P^2}) \right|
_{\phi =
\langle
\phi \rangle}~.
\label{33}
\eea
Here $\sqrt{4\pi G} m^2_i$ is the same strength as for (static)
gravity,
the
correction $Z^{1/2}_\phi$ is $1 + O(\alpha_{GUT})$, while the last
factor is
computable if $e^\phi$ is reasonably small.  One finds$^{11)}$
 that the dilaton coupling to nuclear matter (QCD confinement mass)
is
about 80 times larger than gravity.  Furthermore, since the coupling
to
electromagnetic mass or to leptons is expected to be substantially
different
(probably smaller), one predicts a composition-dependent ``5th force"
of
strength
larger than gravity.  Tests of the equivalence principle down
 to ranges of $10^{-1}$ - 1~ cm$^{12)}$ thus put  bounds$^{11),13)}$
 on the dilaton mass [for a possible way out see, however,
ref. 14)], i.e.
\beq
  M_\phi > 10^{-4}~{\rm eV}~.
\label{34}
\eeq

{\bf 7.} Details about the dilaton potential are unknown,
yet:
\begin{itemize}
\item[a)] On theoretical grounds, in critical superstring theory, the
dilaton
potential has to go to zero as a double
exponential as $\phi \rightarrow -\infty$ (weak coupling):
\beq
V(\phi) \sim {\rm exp}\left( -c^2{\rm exp}(-\phi) \right) = {\rm exp}
\left( -
\frac{c^2}{4 \pi \alpha_{GUT}} \right)~,
\label{36}
\eeq
with $c^2$ a positive (but model-dependent) constant.
\item[b)] On physical grounds it should have a non-trivial minimum at
its present
 value ($ \langle
\phi \rangle = \phi_0 \sim 0)$ with
a vanishing cosmological constant, $V(\phi_0) = 0$.
\end{itemize}

A typical potential satisfying a) and b) is shown in Fig. 1.  The
dotted
lines at
$\phi > 0$ represent our ignorance about strongly coupled string
theory.
Fortunately, the details of what happens in that region will not be
very
relevant
for our subsequent discussion.

\begin{figure}[H]
\hglue 1.3cm
\epsfig{figure=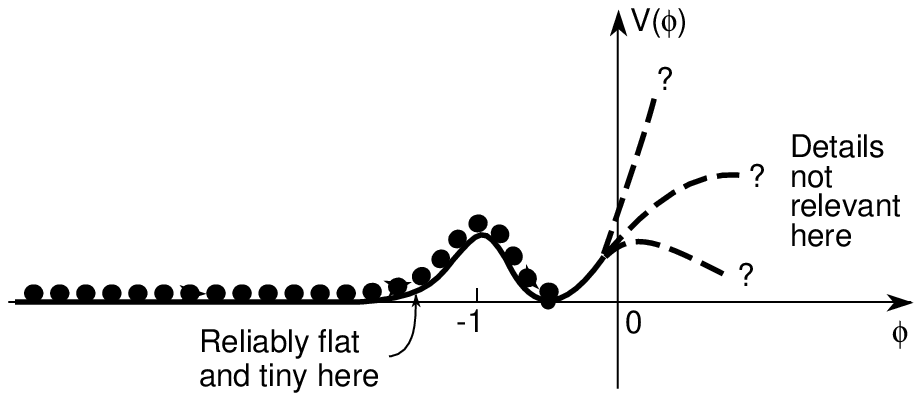,width=13cm}
\caption[]{A possible dilaton potential with illustration of an
inflation-driving rolling dilaton (large dots).
} \end{figure}

{\bf 8.}
There is an exact (all-order) vacuum solution for (critical)
superstring
theory. Unfortunately, it corresponds to a free theory ($g=0$ or
$\phi =
-\infty$) in flat, ten-dimensional, Minkowski space-time, nothing
like
the
world we seem to be living in!

\section{The main ideas/assumptions of string cosmology}
\setcounter{equation}{0}

The very basic postulate of (our own version of) String
Cosmology$^{15),16)}$
 is
that the Universe did indeed start near its trivial vacuum
mentioned at the end of the previous section.

 Fortunately, if one
looks at the space of homogeneous (and for simplicity spacially-flat)
perturbative vacuum solutions, one finds that the trivial vacuum
is a very special, $\it{unstable}$ solution. This is depicted
in Fig.~2a for the simplest case of a ten-dimensional
cosmology in which three spatial dimensions evolve isotropically
while six ``internal" dimensions are static (it is easy to generalize
the discussion to the case of dynamical internal dimensions, but
then the picture becomes multidimensional).

The straight lines in the $H, \dot{\bar{\phi}}$ plane (where
$\dot{\bar{\phi}} \equiv \dot{\phi} - 3 H$) represent
the evolution of the scale factor and of the coupling constant
 as a function of the cosmic time parameter (arrows
along the lines show the direction of the time evolution).
As a consequence of a stringy symmetry,
known$^{15),17)}$ as ``Scale Factor Duality (SFD)",
 there are two branches
(two straight lines). Furthermore, each branch  is
split by the origin in two time-reversal-related parts
(time reversal changes the sign of both
 $H$ and $\dot{\bar{\phi}}$).

The origin (the trivial vacuum) is an ``unstable"
 fixed point: a small perturbation in the direction of positive
$\dot{\bar{\phi}}$ makes the system evolve further and further
from the origin, meaning larger and larger coupling and
absolute value of the Hubble parameter.
This means an accelerated expansion or an accelerated contraction,
 i.e. in the latter case,
inflation. It is tempting to assume that those patches of the
original
Universe that had the right kind of fluctuation  have  grown up
to become (by far) the largest fraction of the Universe today.

In order to arrive at a physically interesting scenario, however,
 we have to connect somehow the top-right inflationary branch
to  the bottom-right branch, since the latter is nothing
but the standard FRW cosmology, which has presumably prevailed
for the last few billion years or so.
Here the so-called ``exit problem" arises. At lowest order
in $\lambda^2$ (small curvatures in string units)
the two branches do
not talk to each other. The inflationary (also called $+$) branch
has a singularity in the future (it takes a finite cosmic
time to reach $\infty$ in our gragh if one starts
from anywhere but  the origin) while the FRW ($-$) branch
has a singularity in the past (the usual big-bang singularity).

It is widely believed that QST has a way to avoid the usual
singularities
of Classical General Relativity or at least a way to reinterpret
them$^{18),19)}$. It thus looks reasonable to assume that the
inflationary
branch, instead of leading to a non-sensical singularity, will
evolve into the FRW branch at values of $\lambda^2$ of order unity.
This is schematically \hfill

\begin{figure}[H]
\hglue 1.3cm
\epsfig{figure=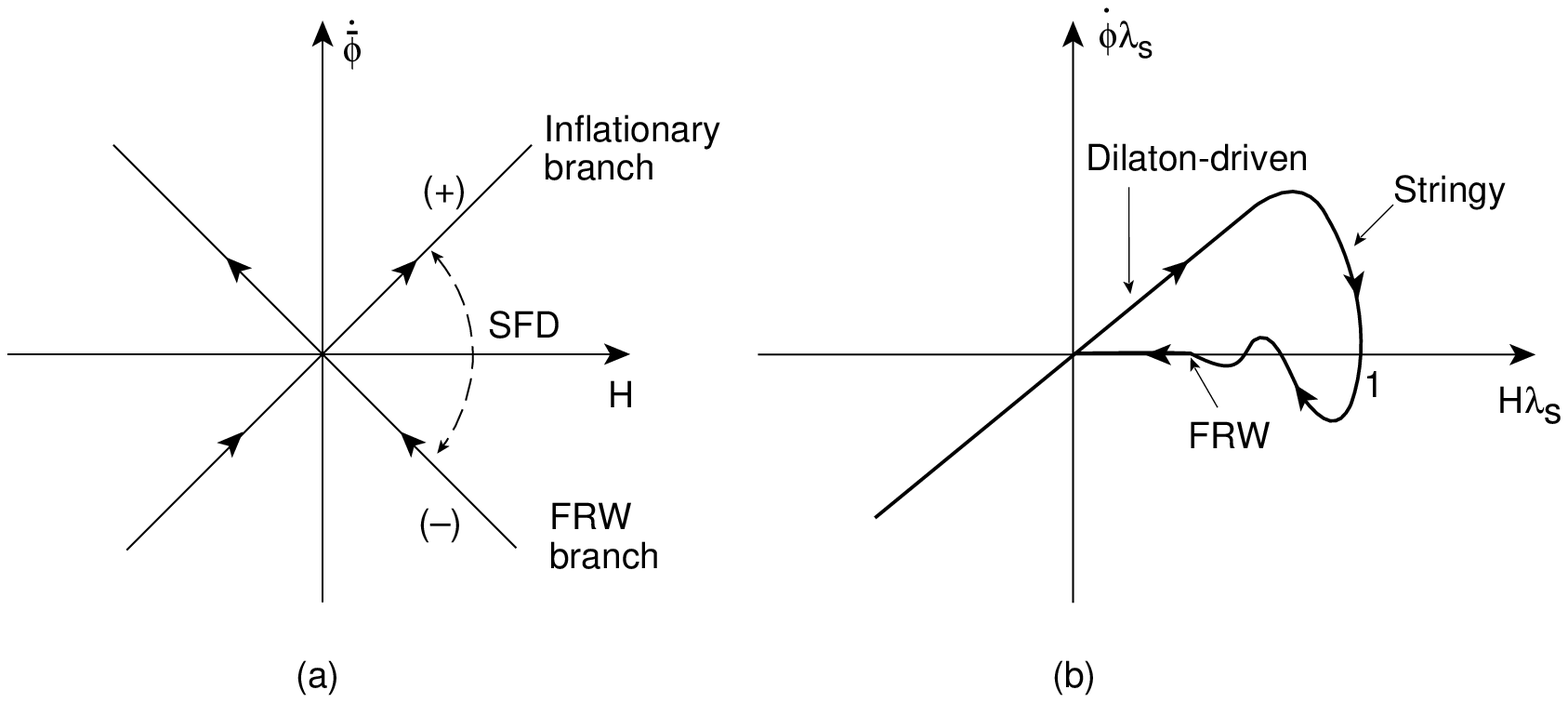,width=13cm}
\caption[]{Phase diagrams for the perturbative (a) regime and a
conjectured non-perturbative solution (b) to the branch-change
problem.
} \end{figure}

\noindent
shown in Fig. 2b, where we have gone back
from $\dot{\bar{\phi}}$ to $\dot{\phi}$
and we have implicitly taken into account the effects of a
non-vanishing dilaton potential at small $\phi$ in order to
freeze the dilaton at its present value.
The need for the branch change to occur at large $\lambda^2$,
first argued for in$^{20)}$, has been recently proved
in ref. 21).

There is a rather simple way to
parametrize a class of scenarios of the kind defined above.
They contain (roughly) three phases and two parameters. Indeed:

In phase I the Universe evolves at $g^2, \lambda^2 \ll 1$
and thus is close to the trivial vacuum. This phase
can  be studied using the tree-level low-energy effective action
(\ref{31}) and is characterized by a long period of dilaton-driven
inflation. The accelerated expansion of the Universe, instead of
originating from the potential
energy of an inflaton field, is
driven by the growth of the coupling constant (i.e. by the dilaton's
kinetic energy, see ref. 22) for a
similar kind of inflationary scenario)
 with $\dot{\phi} = 2\dot{g} / g \sim H$ during the whole phase.

Phase I  supposedly  ends when the coupling $\lambda^2$
reaches values of $O(1)$, so that higher-derivative terms in
the effective action become relevant. Assuming that this happens
 while $g^2$ is still small (and thus the potential is
still negligible), the value $g_s$  of
$g$ at the end of phase I (the beginning
of phase II) is an arbitrary parameter (a modulus of the solution).

During phase II, the stringy version of the big bang,
 the curvature, as well as $\dot\phi$, are
 assumed to remain fixed at their maximal value given by
the string scale (i.e. we expect $\lambda \sim 1$).
The coupling $g$ will instead continue to grow from the value $g_s$
until it is its own turn to
 reach  values $O(1)$. At that point, assuming a
branch change  to have occurred at large curvatures, the dilaton
will be attracted to the true non-perturbative minimum of its
potential; the standard FRW cosmology can then start, provided
the Universe was heated-up  and filled with radiation
(this is not a problem, see below).
The second important parameter of this scenario is the duration of
phase
II
or better the total red-shift, $z_s \equiv a_{end}/a_{beg}$,
which has occurred from the beginning to the end of the stringy
phase.

Our present ignorance about this most crucial phase (and in
particular
about the way the exit can be implemented) prevents us from having a
better description of this phase which, in principle, should not
introduce  new arbitrary parameters ($z_s$ should be eventually
determined
in terms of $g_s$).

During Phase III, the Universe evolves towards smaller and smaller
curvatures but stays at moderate-to-strong coupling. This is the
regime
in
which usual QFT methods are applicable. The details of the
particular gauge theory emerging from the string's non-perturbative
vacuum will be very important in determining the subsequent evolution
and in particular the problem of structure formation, dark matter
and the like.

 Our scenario contains implicitly an arrow
of time, which points in the direction of increasing entropy,
inhomogeneity and structure. As a result of the amplification of
primordial vacuum fluctuations, the Universe is not coming
back to its initial simple (and unique) state (the origin in
Fig. 2), but to the much more structured (and interesting)
state in which we are living today. Actually, the arrow of time
itself should be determined by the direction in which entropy
(and complexity) are growing. This will force us
to identify ($\it{by~definition}$) the perturbative
vacuum  with the initial state of the Universe!

\section{Observable consequences}
\setcounter{equation}{0}

All the observable consequences I will discuss below have
something to do with the well-known phenomenon$^{23)}$ of
amplification of vacuum quantum fluctuations in
cosmological backgrounds. Any conformally flat cosmological
background is known:
\begin{itemize}
\item[a)] to amplify tensor perturbations, i.e. to produce
a stochastic background of gravitational waves;
\item[b)] to induce scalar-metric perturbations from the coupling
of the metric either to a fluid or to scalar particles (in our
context
to the dilaton).

\end{itemize}
By contrast, because of the scale-invariant coupling
of gauge fields in four dimensions, electromagnetic (EM)
perturbations are $\it{not}$ amplified in a conformally flat
cosmological background (even if inflationary).
In string cosmology, the presence of a time-dependent dilaton
in front of the gauge-field kinetic term yields, on top of the
two previously mentioned effects,
\begin{itemize}
\item[c)] an amplification of EM perturbations
corresponding to the creation of macroscopic magnetic (and electric)
fields.
\end{itemize}

Various physically interesting questions arise in connection with
the three effects I have just mentioned. These include the following:
\begin{itemize}
\item[1.] Does the Universe remain quasi-homogeneous during the whole
string-cosmology history?
\item[2.] Does one generate a phenomenologically interesting (i.e.
measurable) background of GW?
\item[3.] Can one produce large enough seeds for generating the
observed galactic (and extragalactic) magnetic fields?
\item[4.] Can scalar, tensor (and possibly EM) perturbations explain
the large-scale anisotropy of the CMB observed by COBE?
\item[5.] Do these perturbations have anything to do with the CMB
itself?
\end{itemize}

In the rest of this talk I will first explain, on the toy example of
the harmonic oscillator, the common mechanism by which quantum
fluctuations are amplified in cosmological backgrounds. I will then
give our present answers to the questions listed above, leaving
details and derivations to the talk of M. Gasperini$^{24)}$.

Consider a one-dimensional (non-relativistic)
 harmonic oscillator moving in a
cosmological background of the simplest kind, characterized by
a scale factor $a(t)$. In units in which  the mass of the
oscillator is $1$, the Lagrangian reads:
\beq
L = {1 \over 2} a^2 (\dot{x}^2 - \omega^2 x^2)
\label{lagra}
\eeq
while the canonical momentum and Hamiltonian are given by
\beq
p = a^2  \dot{x}\; , \;\quad H = {1 \over 2} (a^{-2} p^2  + a^2
\omega^2
x^2) ~.
\label{ham}
\eeq
Let us first discuss the solutions of the classical equations
of motion:
\beq
\ddot{x} + 2~ \dot{a}/a ~\dot{x} + \omega^2 x = 0~, \quad\quad
\ddot{y}  + (\omega^2 - \ddot{a}/a) y = 0 \; ,
\label{motion}
\eeq
where $y \equiv a x$ is the proper (physical) amplitude as opposed
to the comoving amplitude $x$.

Solutions to Eqs. (\ref{motion}) simplify in two
opposite regimes:

a) For $\omega^2 \gg \ddot{a}/a$ there is ``adiabatic damping"
of the comoving amplitude (the name is clearly unappropriate
in the case of contraction):
\beq
x \sim a^{-1}~ e^{\pm i \omega t}~, \quad\quad
p \sim a ~ \omega ~ e^{\pm i \omega t}
\label{damping}
\eeq
which means that, in this regime,  the proper amplitude $y$
and the proper momentum $p/a$ stay constant (and so does the
Hamiltonian).

b) For $\omega^2 \ll \ddot{a}/a$ one finds the so-called
``freeze-out" regime in which:
\bea
x \sim  B ~+~ C \int_0^t dt' a^{-2}(t')\nonumber \\
p \sim  C + \dots
\label{freeze}
\eea
where the comoving amplitude and momentum are fixed.
In this regime the Hamiltonian (the energy) of the system tends to
grow at late times
 whenever $a$ increases or decreases by a large
factor during the freeze-out regime. In the former case the energy
is dominated asymptotically
by the term proportional to $x^2$ and is due to
the ``stretching" of the oscillator caused by the fast expansion,
 while
in the latter case the term proportional to $p^2$ dominates
 because of
the large  blue-shift suffered by the momentum in a contracting
background.

Consider now a cosmology such that
\bea
\omega^2 > \ddot{a}/a \, , ~~~~~\, t < t_{ex} , \, t >
t_{re}\nonumber
\\
\omega^2 < \ddot{a}/a \, , \, ~~~~ t_{ex} < t < t_{re}
\eea
where, anticipating our subsequent discussion, we have defined
the moments of exit and re-entry by the condition $\omega \sim H$.
Such an example will be typical of our scenario, since a given scale
will be well inside the horizon at the beginning (small Hubble
parameter), outside during the high-curvature regime, and then
inside again after re-entry. By joining smoothly the two asymptotic
solutions, we easily find that
the energy of the harmonic oscillator
(which is constant during the
initial and final phases) has been amplified during the
intermediate phase by a factor:
\beq
|c|^2 = Max\left({a^2(t_{re})\over a^2(t_{ex})}~~,~~
{a^2(t_{ex}) \over a^2(t_{re})}\right)
\label{bog}
\eeq
corresponding to the two above-mentioned cases.

The excercise can be repeated  at the quantum level starting, for
instance, from a harmonic oscillator in its ground state.
Quantum mechanics fixes the size of the initial amplitude,
momentum and energy:
\beq
|x| \sim a^{-1} \sqrt{{\hbar \over \omega}} \, , ~~
 E \sim \hbar \omega~.
\eeq

The quantum mechanical interpretation of eq. (\ref{bog}) is
that $c$ is the Bogoliubov coefficient transforming the initial
ground state into the final excited quantum state
($|c|^2$ being the average occupation number for the
latter). Note that the final state
 ends up being highly ``squeezed", i.e. having a large $\Delta x$
or $\Delta p$ depending on the sign of $H$. If, because of
coarse-graining, the squeezed coordinate is not measured,
the final state will look like a high-entropy, statistical
ensemble of quasi-classical oscillators.

Up to technical complications, things work out pretty much in the
same
way for strings$^{25)}$ and for the three kinds of perturbations
mentioned at the beginning of this section. In particular, for
each one of the latter, one can define$^{26)}$
a canonical variable
$\psi^{i}$ (similar
to the harmonic oscillator's $y$) satisfying an
equation of the type
\beq
\psi^{i}_{k} + [k^2 - V_{i}(\eta)] \psi^{i}_{k} = 0~,
\label{perteq}
\eeq
where $i$ labels the type of perturbation, $k$ is the comoving
wave number and $\eta$ is the conformal time (derivatives with
respect
to which
 are denoted by a prime).

Since, for each $i$, the ``potential" $V_i$ is very small
at very early times,  grows to a maximum during the stringy era and,
finally, drops rapidly to zero at the beginning of the radiation era,
a given scale ($k$) begins and ends inside the horizon with an
intermediate phase outside. Larger scales exit first and re-enter
later. Also, in our scenario,
 larger scales exit and re-enter at smaller values of $H$.
 Very short scales
exit during the stringy era and, for those, our predictions will
not be as solid as for the scales that leave the horizon
during the perturbative dilatonic phase I. The fact that
the amplification of perturbation depends just on some
ratios of fields evaluated at exit and re-entry (and not on the
details of the evolution in between) makes us believe that our
detailed results are trustworthy for those larger scales.
This being said, I  present below
 some results on the five issues mentioned above (see, again,
ref. 24) for derivations and/or
details).

\begin{itemize}
\item[1.] {\bf Does the Universe remain quasi-homogeneous during the
whole
string-cosmology history?}

\vspace*{0.2cm}
The answer to this question turns out to be yes! This
is not a priori evident since,
in commonly used gauges$^{26)}$ for
scalar perturbations of the metric (e.g. the so-called longitudinal
gauge in which the metric remains diagonal),
 such perturbations appear
to grow very large during the inflationary phase and to destroy
homogeneity or, at least, to prevent the use of linear perturbation
theory. Similar problems had been encountered earlier in the context
of Kaluza-Klein cosmology$^{27)}$.

In ref. 28) it was shown that, by a suitable choice of gauge
(an ``off-diagonal" gauge), the growing mode of the perturbation
can be tamed. This can be double-checked by using the so-called
gauge-invariant variables of Bruni and Ellis$^{29)}$.
The bottom line is that scalar perturbations in string cosmology
behave no worse than tensor perturbations,
 to which we now turn our attention.

\item[2.] {\bf Does one generate a phenomenologically interesting
(i.e.
measurable) background of GW?}

\vspace*{0.2cm}
The canonical variable $\psi$
for tensor perturbations (i.e. for GW) is defined by:
\bea
g_{\mu\nu} = a^2 [\eta_{\mu\nu} + h_{\mu\nu} ] \nonumber \\
\psi = (a/g) ~ h = a ~ e^{-\phi/2} h~,
\eea
where $h$ stands for either  of the two transverse-traceless
polarizations of the gravitational wave.
As long as the perturbation is inside the horizon,
$\psi$ remains constant while $h$ is adiabatically damped. By
contrast,
 outside the horizon, $\psi$ is amplified according to
\beq
\psi_k \sim (a/g)[C_k + D_k \int_{\eta_{ex}}^{\eta}~
d \eta ' ~ g^2(\eta ') ~ a^{-2}(\eta ')]
\label{tensorsol}
\eeq
where, for each Fourier mode of (comoving) wave number $k$,
$\eta_{ex}= k^{-1}$.

The first term in (\ref{tensorsol}) clearly
corresponds to the freezing of
$h$ itself, while the second term represents the freezing of its
associated canonical momentum. In standard (non-dilatonic)
inflationary models, the first term dominates since
$a$ grows very fast. In our case,  the second term dominates
since the growth of $a$ is
 over-compensated
by the growth of $g$ (i.e. of $\phi$).
 This is equivalent to
saying that, in the Einstein frame, our background describes
 a contracting Universe.

After matching the result (\ref{tensorsol}) with the
usual oscillatory, damped behaviour of the radiation-dominated
epoch, one arrives at the final result$^{28),30)}$
 for the magnitude
of the stochastic background of GW today:
\bea
|\delta h_{\omega}|  \equiv k^{3/2} |h_k| &\sim &
\sqrt{{H_0 \over M_s}} z_{eq}^{-1/4} z_s \left({g_s \over g_1}\right)
\left({\omega \over \omega_s}\right)^{1/2}~ \nonumber \\
&& \left[ \ln
\left({\omega_s \over \omega}\right)
+  (z_s)^{-3} \left({g_s \over g_1}\right)^{-2} ~\right]~,\;
 \quad\quad \omega < \omega_s \hfill
\label{deltah}
\eea
where $\omega = k/a$ is the proper frequency, $z_{eq} \sim 10^4,
\omega_s \sim z_s^{-1} (g_1)^{1/2} \times 10^{11}$ Hz $\equiv
z_s^{-1} \omega_1$.

The above result can be converted into a spectrum of energy
density per logarithmic interval of frequency. In critical
density units:
\beq
{d \Omega_{GW} \over d \ln \omega} = z_{eq}^{-1}(g_s)^2
\left({\omega \over \omega_s}\right)^3~ \left[ \ln
\left({\omega_s \over \omega}\right)
+  (z_s)^{-3} \left({g_s \over g_1}\right)^{-2} ~\right]^2 ~ ,\;
\omega < \omega_s~.
\label{Omegagw}
\eeq
The above spectrum looks quasi-thermal al large scales (i.e.
at $\omega < \omega_s$), but is amplified by a large factor
relative to a Planckian spectrum of temperature $\omega_s$.
The spectrum is expected to extend above $\omega_s$
 up to $\omega_1$, but the shape and magnitude of the spectrum in
this interval are too dependent on the details of the stringy
phase to be really trustworthy.

In Fig. 3 we show the spectrum of stochastic gravitational
waves expected from our two-parameter model. For a given pair
$g_s, z_s$ one identifies a point in the $\omega, \delta h$ plane
that represents the end-point $\omega_s, \delta h_{\omega_s}$
of the $\omega^{1/2}$ spectrum,  \hfill

\begin{figure}[H]
\hglue 1.3cm
\epsfig{figure=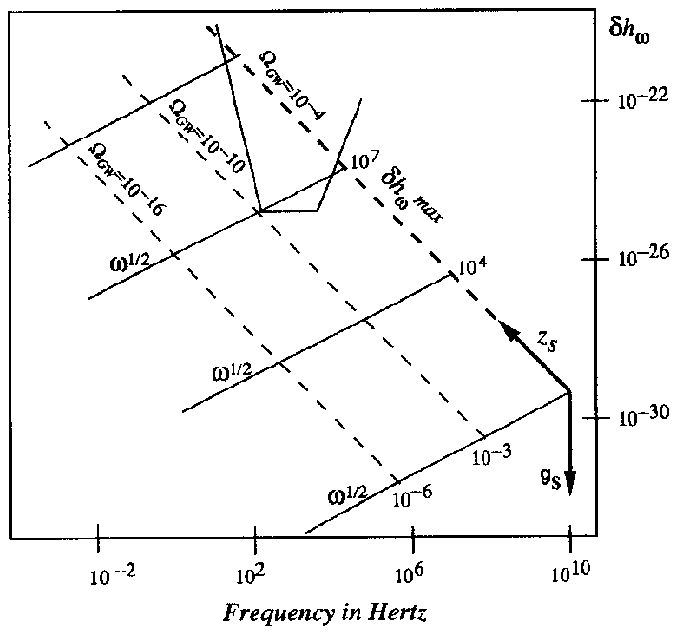,width=13cm}
\caption[]{GW spectra from string cosmology against
interferometric sensitivity.
} \end{figure}

\noindent
which holds for scales
crossing the horizon during the dilatonic era
(the rest of the spectrum,
which is more model-dependent, is not shown).
The odd-shaped region in Fig. 3 shows
the expected sensitivity of the so-called ``Advanced LIGO"
project$^{31)}$. Resonant bars$^{32)}$ might also be able to reach
comparable sensitivity in the kHz region, while microwave
cavities, if conveniently developed, could be used in the region
$10^6$-$10^9$ Hz $^{33)}$.
 Another interesting possibility
is coincidence experiments between an interferometer and a bar.
 The quoted sensitivity$^{34)}$ to a stochastic background, as a
 function of the frequency $f$,
of the individual sensitivities of the bar and of the interferometer
 $\delta h$,
and of the observation time $T_{obs}$,
is:
\beq
\delta\Omega_{GW} = 1.5 ~ 10^{-5} (f/10^3{\rm Hz})^3 (10^{21} \delta
h_{int})
(10^{21} \delta h_{bar})(T_{obs}/10^7 ~{\rm s})^{-1/2}.
\label{coinc}
\eeq

Obviously, detecting a stochastic backgound like ours
 is a formidable challenge. Also, the physical range of our
parameters $g_s, z_s$ could be such that no observable
signal will be produced. What is perhaps more interesting
is the fact that
there are  cosmological models that
 predict a non-negligible
yield of GW in a range of
 frequencies where other sources predict just a ``desert".
We hope that this will encourage experimentalists to
develop suitable techniques to reach sensitivities
better than $\delta h = 10^{-24}$ in the $10^6$-$10^9~{\rm  Hz}$
region.

\item[3.] {\bf Can   large enough seeds be produced for
the generation of
observed galactic (and extragalactic) magnetic fields?}

\vspace*{0.2cm}
As already mentioned, seeds for generating
the galactic magnetic fields through the so-called cosmic dynamo
mechanism$^{35)}$
 can be generated in our scenario
 by the amplification of the quantum  fluctuations of the EM field.
In this case the canonical variable is just the
(Fourier transform of the) usual $A_{\mu}$ potential.
In analogy with (\ref{tensorsol}) its
 amplification, while outside the horizon,
 is described by the asymptotic solution:
\beq
A_k \sim g^{-1}\left[C_k + D_k \int_{\eta_{ex}}^{\eta}~
d \eta ' g^2 (\eta ') \right]~,
\label{elsol}
\eeq
which leads$^{36),37)}$ to an overall amplification
of the electromagnetic field
by a factor $|c_{k}|^2 \sim (g_{re}/g_{ex})^2$.
Again, the main contribution to the amplification comes from
the second term on the r.h.s. of eq. (\ref{elsol}).
One can express this result in terms of the
 fraction of electromagnetic
energy stored in a unit of logarithmic interval of $\omega$
normalized to the one in the CMB, $\rho_{\gamma}$. One finds:
\beq
r(\omega)=\frac{\omega}{\rho_{\gamma}}
 \frac{d\rho_{B}}{d\omega} \simeq
\frac{\omega ^{4}}{\rho_{\gamma}} |c_{-}(\omega)|^2 \equiv
\frac{\omega^{4}}{\rho_{\gamma}}(g_{re}/g_{ex})^2~.
\label{r}
\end{equation}

The ratio $r(\omega)$ stays
constant
 during the phase of matter-dominated as well as radiation-dominated
evolution, in which the Universe behaves like a good electromagnetic
 conductor$^{38)}$.
In terms of $r(\omega)$ the condition that has to be satisfied in
order
to
seed in the galaxies a magnetic field big enough to be amplified by
the
ordinary mechanisms of plasma physics is$^{38)}$
\beq
r(\omega_{G})\geq 10^{-34}
\label{condition}
\eeq
where  $\omega_{G}\simeq (1$
Mpc$)^{-1}\simeq 10^{-14}$ Hz is the galactic scale.
Using the known value of $\rho_{\gamma}$, we thus find, from
(\ref{r},
\ref{condition}):
\begin{equation}
 g_{ex} < 10^{-33}~,
\label{rr}
\end{equation}
i.e. a very tiny coupling at the time of exit
 of the galactic scale.

The conclusion is that string (or more generally dilaton-driven)
inflation stands a unique chance of explaining the origin of
the galactic magnetic fields. Indeed, if the seeds of the magnetic
fields
are to be attributed to the amplification of vacuum fluctuations,
their present magnitude can be interpreted as evidence that the
fine structure constant has evolved to its present value
from a tiny one during inflation.

\item[4.] {\bf Can scalar, tensor (and possibly EM)
perturbations explain
the large-scale anisotropy of the CMB observed by COBE?}

\vspace*{0.2cm}
The answer here is certainly negative as far as scalar and
tensor perturbations are concerned. The reason is simple:
for spectra that are normalized to $O(1)$ (at most)
at the maximal amplified frequency $\omega_1 \sim 10^{11}$ Hz,
 and that grow  like $\omega^{1/2}$, one
cannot have any substantial power at the scales $O(10^{-18}$Hz)
to which COBE is sensitive. The origin of $\Delta T / T$ at large
scale
would have to be attributed to other effects
 (e.g. topological defects).

There is however a (small?) chance$^{39)}$ that the EM
perturbations themselves might explain the anisotropies of the CMB
since the spectrum of electromagnetic perturbations turns out
to be flatter (and more model-dependent)
than that of metric perturbations. Assuming that this is the
case,
an interesting relation is obtained$^{39)}$ between the magnitute
of large scale anysotropies and the slope of the power spectrum.
Such a relation turns out to be fully consistent, with present
bounds on the spectral index.

\item[5.] {\bf Do all these perturbations have anything
 to do with the CMB itself?}

\vspace*{0.2cm}
Stated differently, this is the question of how to arrive at the
hot big bang of the SCM starting from our
``cold" initial conditions.
The reason why a hot universe can emerge at the end of our
inflationary epochs (phases I and II)
 goes back to an idea of L. Parker$^{40)}$,
 according to which  amplified quantum
flluctuations can
give origin to the CMB itself if Planckian scales are reached.

Rephrasing Parker's idea in our context amounts to solving the
following bootstrap-like condition: at which moment, if any,
 will the energy stored in the perturbations reach
 the critical density?

The total energy density
$\rho_{qf}$ stored
in the amplified vacuum quantum fluctuations is given by:
\beq
\rho_{qf} \sim N_{eff}~ {M_s^4 \over 4 \pi^2}
\left( a_1 / a \right)^4~,
\label{rhoqf}
\eeq
where $N_{eff}$ is the number of effective (relativistic) species,
which get produced (whose energy density decreases like $a^{-4}$)
and $a_1$ is the scale factor at the (supposed) moment of
branch-change.
The critical density (in the same units) is given by:
\beq
\rho_{cr} = e^{-\phi} M_s^2 H^2~.
\label{rhocr}
\eeq

At the beginning, with $e^{\phi}\ll 1$, $\rho_{qf} \ll \rho_{cr}$
 but, in the $(-)$ branch
solution, $\rho_{cr}$ decreases faster than $\rho_{qf}$ so that,
at some moment, $\rho_{qf}$ will become the dominant sort
of energy while the dilaton kinetic term will become negligible.
It would be interesting to find out what sort of initial
temperatures for the radiation era
will come out of this assumption.
\end{itemize}

\section{Conclusions}

I want to conclude by listing which are, in my opinion,
 the pluses and
 minuses of the scenario I have advocated:

{\bf The Goodies}
\begin{itemize}
\item Inflation comes naturally, without ad-hoc fields and
fine-tuning: there is even an underlying symmetry yielding
inflationary solutions.
\item Initial conditions are natural, yet a simple universe
would evolve into a rich and complex one.
\item The kinematical problems of the SCM are solved.
\item Perturbations do not grow too fast to spoil homogeneity.
\item An interesting characteristic spectrum of GW is generated.
\item Larger-than-usual electromagnetic perturbations are
easily generated and could explain the galactic magnetic fields.
\item A hot big bang could be a natural outcome of our inflationary
scenario.
\end{itemize}

{\bf The Baddies}

\begin{itemize}

\item A scale-invariant spectrum is all but automatic (unlike what
happens
in normal vacuum-energy-driven inflation).
\item Our understanding of the high curvature (stringy) phase
and of the (crucial and necessary) change of branch is still
poorly understood (in spite of recent progress in Conformal Field
Theory).
\end{itemize}
On the whole this does not look like a bad score
 for a four-year-old kid!

\vspace*{1cm}
\noindent
{\bf Acknowledgements}

I would like to acknowledge the help and encouragement of my
collaborators
in the work reported here:  Ramy Brustein (CERN--Beer Sheva),
 Maurizio Gasperini (Turin), Massimo
Giovannini (CERN--Turin), and Slava Mukhanov (Zurich--Moscow).
This work has also benefited from earlier collaborations with
Jnan Maharana (Bhubaneswar), Kris Meissner (Trieste--Varsaw),
Roberto Ricci (CERN--Rome), Norma Sanchez (Paris)
 and Nguyen Suan Han (Hanoi).

  \end{document}